\documentclass[journal=jacsat,manuscript=article]{achemso}

\usepackage[version=3]{mhchem} 

\usepackage{subcaption}
\usepackage{graphicx}
\usepackage{siunitx}

\usepackage{xcolor}

\author{Asmita S. Thool}

\phone{}
\fax{}
\affiliation
{Department of Electrical Engineering, Indian Institute of Technology Madras, Chennai, India}
\author{Sourodeep Roy}
\affiliation
{Department of Electrical Engineering, Indian Institute of Technology Madras, Chennai, India}
\author{Prahalad Kanti Barman}
\affiliation
{Department of Physics, Indian Institute of Technology Madras, Chennai, India}
\author{Kartick Biswas}
\affiliation
{Center for Nano Science and Engineering, Indian Institute of Science, Bangalore, India}
\author{Pavan Nukala}
\affiliation
{Center for Nano Science and Engineering, Indian Institute of Science, Bangalore, India}
\author{Abhishek Misra}
\affiliation
{Department of Physics, Indian Institute of Technology Madras, Chennai, India} 
\author{Saptarshi Das}
\affiliation
{The Pennsylvania State University, University Park, PA, USA} 
\author{Bhaswar Chakrabarti}
\email{bchakrabarti@ee.iitm.ac.in,  
sud70@psu.edu}
\affiliation
{Department of Electrical Engineering, Indian Institute of Technology Madras, Chennai, India}

\title[An \textsf{achemso} demo]
  {Interfacial and bulk switching MoS$_2$ memristors for an all-2D reservoir computing framework}

\abbreviations{IR,NMR,UV}

\begin{document}
\begin{abstract}
In this study, we design a reservoir computing (RC) network by exploiting short- and long-term memory dynamics in  Au/Ti/MoS$_2$/Au memristive devices. The temporal dynamics is engineered by controlling the thickness of the Chemical Vapor Deposited (CVD) MoS$_2$ films. Devices with a monolayer (1L)-MoS$_2$ film exhibit volatile (short-term memory) switching dynamics. We also report non-volatile resistance switching with excellent uniformity and analog behavior in conductance tuning for the multilayer (ML) MoS$_2$ memristive devices. We correlate this performance with trap-assisted space-charge limited conduction (SCLC) mechanism, leading to a bulk-limited resistance switching behavior. Four-bit reservoir states are generated using volatile memristors. The readout layer is implemented with an array of nonvolatile synapses. This small RC network achieves 89.56\% precision in a spoken-digit recognition task and is also used to analyze a nonlinear time series equation.\\
\textbf{\(Keywords\)} - Monolayer (1L)-MoS$_2$, Multilayer (ML)-MoS$_2$,  Short Term Memory (STM), Long Term Memory (LTM), Reservoir Computing (RC), Long Term Potentiation (LTP), Long-term Depression (LTD)
\end{abstract}
\section{Introduction}
Despite the enormous success of artificial intelligence (AI) and machine learning, the operation of modern deep neural networks (DNNs) has always required significant computational resources. With increasing complexity of tasks, the size of networks and the volume of training data have increased rapidly \cite{R_1}. The storage and computational requirements of state-of-the-art machine learning frameworks have reached such a level that the sustainability of AI/machine learning is becoming a concern \cite{R_2,R_3}.

The concept of using dynamical systems as reservoirs for converting temporal information into higher-dimensional data has existed for the last two decades\cite{R_4, R_5, R_6, R_7, R_8}. Since then, reservoir computing (RC) has developed considerably as a ML framework. RC networks use a reservoir layer consisting of neurons that evolve dynamically with time and map input data to reservoir states\cite{R_9, R_10}. The reservoir states are analyzed in a readout layer. There are two significant advantages of this approach over conventional DNNs. Firstly, the reservoir layer eliminates the need for multiple hidden layers, thereby reducing the complexity of the network. Secondly, the connections in the reservoir are always fixed and only the weights in the readout layer are adjusted during training \cite{R_10}. As a result, RC can be a significantly more energy-efficient alternative to deep recurrent neural networks (RNN) for analyzing complex dynamical systems.\\ 
The primary challenge in any hardware implementation of RC is to mimic the dynamic behavior of the reservoir neurons. Specifically, a reservoir neuron must possess short-term memory to "remember" the influence of inputs from past time steps\cite{R_9}. Some early attempts have used custom-made application specific integrated circuits (ASIC)\cite{R_11} or field programmable gate arrays (FPGA)\cite{R_12}. For example, Verstraeten et al. demonstrated RC using stochastic bit stream neurons implemented in FPGAs.\cite{R_12} Sch{\"u}rmann et al. have implemented a Liquid State Machine using a general-purpose Perceptron network with recurrent connections.\cite{R_11} However, such implementations typically require complex circuitry due to the lack of short-term memory dynamics in digital technologies. However, some emerging devices naturally exhibit this temporal behavior under certain operating conditions. Torrejon et al. have exploited the dynamic characteristics of spintronic oscillators for this purpose\cite{R_13}. Diffusive memristors are also known to display short-term memory characteristics \cite{R_14, R_15}. Consequently, several recent works have exploited this behavior to design memristive RC frameworks \cite{R_15, R_24, R_25, R_27, R_28}. For example, Du et al. have used an array of WO$_x$ diffusive memristors to build an RC system \cite{R_15}. However, practical implementation of memristive RC, just like other memristive neural networks can suffer from the reliability issues of memristive devices. More specifically, the inherent stochasticity of filamentary memristors can be challenging for accurate programming of device conductance. Complicated conductance tuning algorithms and/or additional circuitry is often required to address these issues, resulting in latency, energy and area overheads. As a result, there has been a growing interest to develop non-filamentary memristors\cite{R_34, R_35, R_36}. However, implementation of an entire neural network using bulk or non-filamentary memristors has been rarely achieved \cite{R_34}. In the context of RC, no prior work has demonstrated a memristive RC network using non-filamentary devices. \\
In this work, we demonstrate reservoir computing using a fully memristive framework by exploiting non-filamentary resistance switching in Au/Ti/MoS$_2$/Au memristive devices. Our original contributions are outlined below. Firstly, we develop non-filamentary MoS$_2$ memristors using large-area, Chemical Vapor Deposited (CVD), 2D-MoS$_2$ films. In our prior work, we have demonstrated highly uniform, volatile resistance switching due to Schottky barrier modulation in monolayer (1L) MoS$_2$ films \cite{R_32}. Here, we demonstrate non-volatile analog resistance switching with outstanding uniformity in Au/Ti/ML-MoS$_2$/Au memristors. Secondly, we systematically study the switching mechanism of the ML-MoS$_2$ devices using electrical characterizations and High-Resolution Scanning Transmission Electron Microscopy (HRSTEM). Our studies reveal trap-assisted SCLC to be the dominant mechanism. Through HRSTEM, we correlate this with the generation of S-vacancies and Au intercalates in the bulk of the multilayer MoS$_2$ film during an initial electroforming process. Next, we develop and characterize 16$\times$16 memristive crossbar arrays with ML-MoS$_2$ as the active material. The devices in the crossbar array exhibit long-term analog conductance tunability with less than 4$\%$ cycle-to-cycle (C2C) standard error. Next, we study the short-term memory effects in 1L-MoS$_2$ devices and create 4-bit reservoir states. Finally, an RC network is designed and simulated using the volatile 1L-MoS$_2$ memristors as reservoir neurons and the 16$\times$16 array of ML-MoS$_2$ memristors as the analog synapses in the read-out layer. We then demonstrate spoken-digit classification with 89.56$\%$ accuracy. We also apply the memristive RC network for the prediction of a non-linear dynamic time-series equation. Our work provides a pathway for creating RC frameworks using non-filamentary memristor technologies developed on a single 2D-material platform. 

\section{Results and discussion}
\subsection{Materials Characterization} 
CVD-grown, large-area, monolayer (1L) and multilayer (ML) MoS$_2$ films are characterized in pristine condition (before device fabrication) by Optical microscopy, Atomic Force Microscopy (AFM), Scanning Electron Microscopy (SEM), Raman spectroscopy, and photoluminescence (PL) measurements. Figures 1 (a) and 1 (f) show optical images of the continuous 1L and ML films grown on the SiO$_2$/Si substrates by CVD. The synthesis process has been described in the experimental section. A schematic of the single-zone CVD furnance used for MoS$_2$ growth is given in the Supplementary \textcolor{blue}{Figure S1}. The temperature profile for the growth of monolayer and multilayer MoS$_2$ can be seen in \textcolor{blue}{Figure S2}. 
The thickness of the MoS$_2$ films grown is verified with the help of AFM. Figures 1 (b-c) show the topography and the corresponding height profile (0.8 nm) of the 1L-MoS$_2$ film as observed by AFM, confirming the thickness to be an atomic layer\cite{R_29}. Figure 1(g-h) shows the topography and corresponding height profile of the multilayer MoS$_2$ film. The thickness of the ML-MoS$_2$ film is confirmed to be $\simeq$ 11 nm, which corresponds to 14 layers. These 1L and ML-MoS$_2$ films are used in all subsequent experiments. 
Next, a monochromatic laser source of wavelength 532 nm is used to perform Raman measurements, as shown in Figure 1(d,i). Here, two characteristic peaks, in-plane E$^1$$_2$$_g$ and out-plane A$_1$$_g$ appear at 384.5 cm$^-$$^1$ and 405 cm$^-$$^1$ for monolayer and at 382.6 cm$^-$$^1$ and  409 cm$^-$$^1$ for multilayer MoS$_2$, respectively. The separation between the two peaks is $\approx$ 20.5 cm$^-$$^1$ and 26.4 cm$^-$$^1$, again confirming the thicknesses of the 1L-MoS$_2$ \cite{R_30} and multilayer MoS$_2$ films. PL measurements as shown in Figure 1(e, j) mark the presence of two characteristic excitonic emissions, ie AX$^0$ and BX$^0$ occurring at 1.85 and 2.0 eV respectively for monolayer and at 1.78 and 1.94 eV respectively for multilayer films.  The ratio of AX$^0$/ BX$^0$ is $\approx$ 0.925 and 0.9175 which again confirms the good quality of the samples\cite{R_31}. To study the uniformity of the grown sample, we performed PL and Raman mapping.
Figures 1 (k) and 1 (l) show the intensity distribution of AX$^0$ over a region of the sample.
The distribution of the peak of E$^1$$_2$$_g$ is shown in Figures 1 (m) and 1 (n). 
In order to further investigate the surface morphology, continuity and thickness we have performed SEM analysis as shown in \textcolor{blue}{Figure S3}. The layer thickness can be roughly judged by the depth of the color, as we observe a contrast between the sample and the substrate. The 1L and ML films show continuous growth with clean surfaces observed in large regions on the substrate.
\begin{figure}
    \centering
    \includegraphics[width=\linewidth]{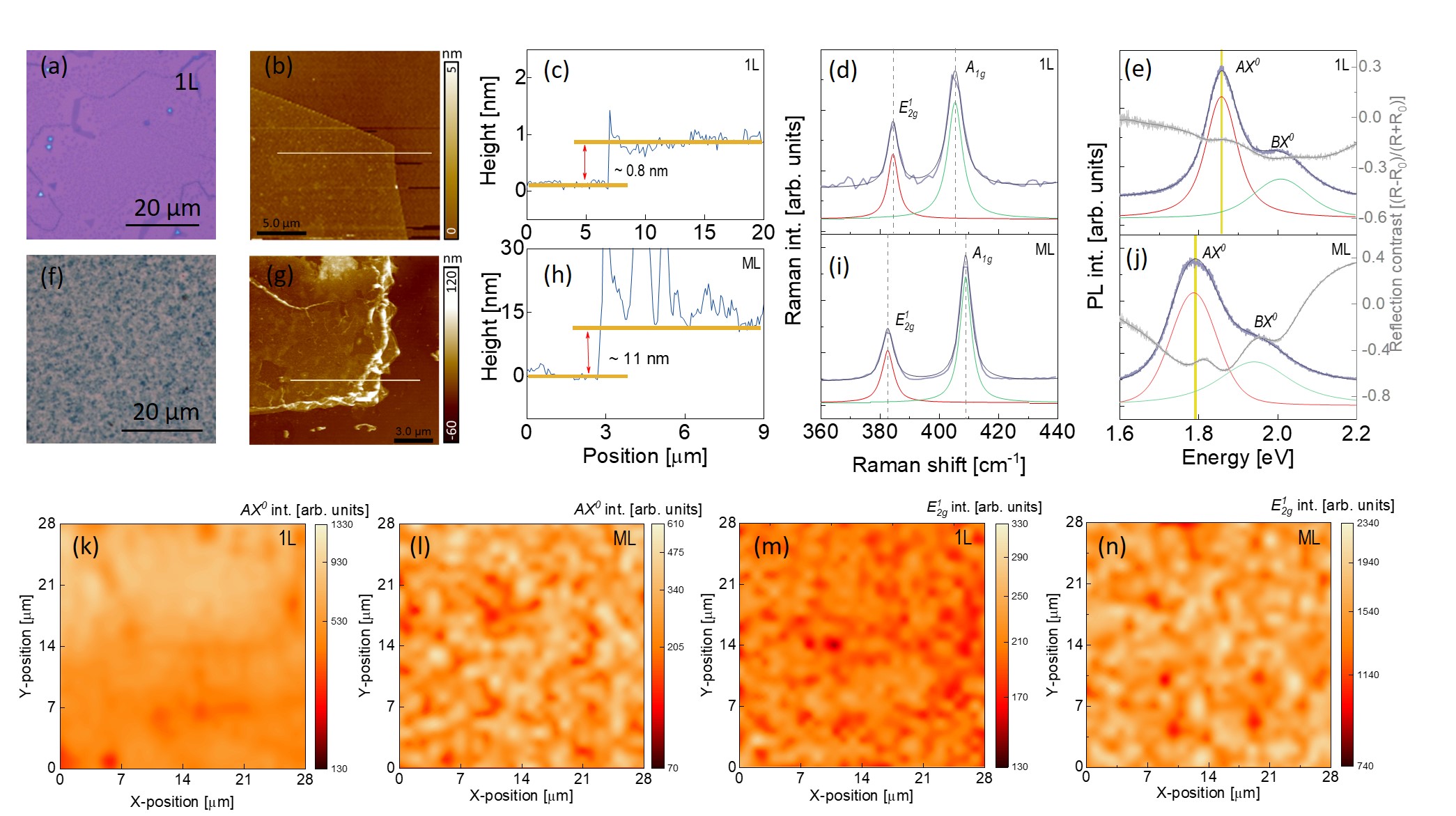}
    \caption{(a,f) Optical image of CVD grown large area continuous film 1L and multilayer MoS$_2$, respectively, (b-c,g-h) AFM image and corresponding height profile of the continuous MoS$_2$ film showing monolayer and multilayer thickness, (d,i) Raman spectroscopy taken on grown MoS$_2$ verifying the growth of 1L and multilayer MoS$_2$, (e,j) Photoluminance spectroscopy taken on grown MoS$_2$ verifying the growth of 1L and multilayer MoS$_2$ (k,l) PL and (m,n) Raman intensity mapping performed on as grown 1L-MoS$_2$ and multilayer MoS$_2$ on the source substrate SiO$_2$/Si.}  
\end{figure}
\subsection{Switching Characteristics and Mechanisms of Transport}
Two terminal memristive crossbar devices and passive crossbar arrays are fabricated on Si/SiO$_2$ substrates with the help of photolithography, metallization, lift-off, chemical transfer process, and etching. Details on the fabrication procedure are given in the experimental section.
Note that the individual devices were fabricated using both 1L-MoS$_2$ or ML-MoS$_2$ as the active material. The fabricated crossbar arrays, on the other hand, use only ML-MoS$_2$ as the active material. We use Au as the bottom electrode and Au/Ti as the top electrode for all our devices. Several prior works, including our own, have noted the effect of electrode materials on the switching characteristics\cite{R_33,R_42, R_43}. It has been observed that Au top and bottom electrodes lead to filament formation due to Au diffusion into MoS$_2$ films. On the other hand, a Ti interfacial layer between MoS$_2$ and Au results analog resistance change due to Schottky barrier modulation at the interface. In addition, Ti provides better adhesion of Au to MoS$_2$ which is crucial for the memristive arrays. We will discuss the switching characteristics of the single devices in this section to understand the differences in behavior between the mono and the multi-layer films and to explore the underlying switching mechanisms.\\
Figure 2(a,b) show the schematic and optical image of a fabricated Au/Ti/1L-MoS$_2$/Au device. Similarly, Figure 2 (c-d) show the schematic and optical image of a Au/Ti/ML-MoS$_2$/Au memristive individual device.  Area of the fabricated 1L and ML-MoS$_2$ devices varies from 2$\times$2 to 10$\times$10 $\mu$m$^2$. For this study, we primarily use 5$\times$5 to 10$\times$10 $\mu$m$^2$ devices.
The STEM cross-sectional image of the whole device structure (Au/Ti/1L-MoS$_2$/Au) on the SiO$_2$/Si substrate, which contains monolayer MoS$_2$ is shown in Figure 2 (e-f) with low magnification.  The monolayer MoS$_2$ stack is shown at a higher magnification in Figure 2(g). The monolayer remained intact and undamaged inside the device.
A similar analysis was performed for one of the multilayer devices fabricated for the study. Figure 2 (h-i) shows the low-magnification images of an ML-MoS$_2$ device containing the structure (Au/Ti/ML-MoS$_2$/Au) on a SiO$_2$/Si substrate. The corresponding EDS mapping of chemical elements is shown in Figure 2(j), which can distinguish the ML-MoS$_2$ and other top and bottom electrode layers and also demonstrate the intrinsic chemical composition of the active layer without noticeable oxidation. A magnified image of the ML device which contains 14 Layer of MoS$_2$ is shown in the Figure 2(k). This STEM image confirms that the MoS$_2$ layers are tightly stacked with vertical geometry.
The periodicity of the layers is found to be 0.62 nm, which is equivalent to the inter-layer separation of MoS$_2$ as shown in Figure 2(l).
A high-quality and well-formed interface between the electrodes and the multilayer film is clearly observed. In addition, the atomically flat 2D interface among the consecutive layers also ensures efficient charge transport across the multilayer stacks, which directly impacts performance in multicycle device operation.
\begin{figure}
    \centering
    \includegraphics[width=\linewidth]{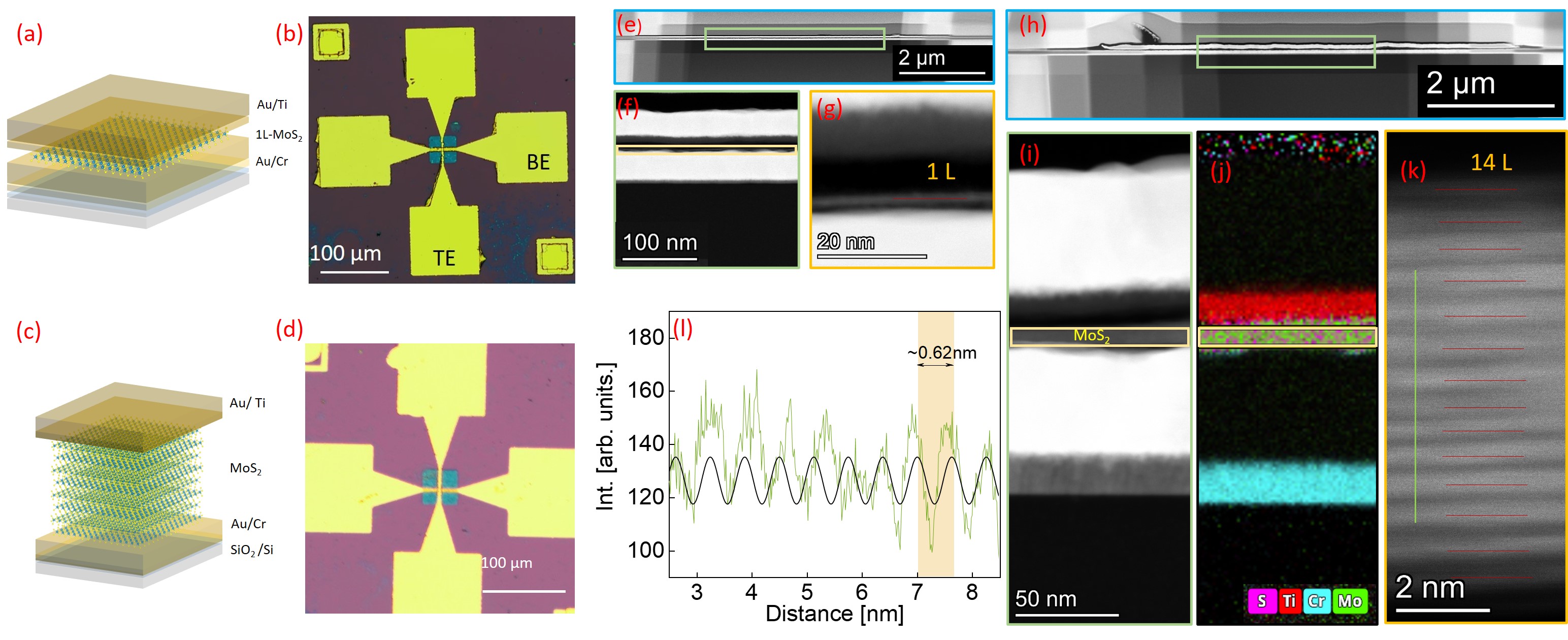}
    \caption{(a) Schematic image of Au/Ti/1L-MoS$_2$/Au device and, its corresponding (b) Optical image of a 5 $\mu$m $\times$ 5 $\mu$m 1L-MoS$_2$ device, (c) Schematic image of Au/Ti/ML-MoS$_2$/Au device and, its corresponding (d) Optical image of a 5 $\mu$m $\times$ 5 $\mu$m ML-MoS$_2$ device, (e-g) HRSTEM image of a 1L MoS$_2$ device, (h-k) HRSTEM images of a ML-MoS$_2$ device, (l) Interlayer periodicity of ML-MoS$_2$ in the fabricated device.}  
\end{figure}
\\The electrical measurements on both types of devices are done by applying the input voltage signals to the bottom electrode while keeping the top electrode at electrical ground. Supplementary \textcolor{blue}{Figure S4(a)} shows the quasi-DC \(I-V\) characteristics of a Au/Ti/1L-MoS$_2$/Au device measured over approximately 100 cycles. The 1L-MoS$_2$ devices do not require a separate electroforming operation. The as-fabricated devices are found in the high-resistance state (HRS). In a bidirectional voltage sweep, the current starts to increase near a threshold voltage (V$_T$) of $\approx$ 1.5 V. However, the positive sweep is further increased to 2.5 V (V$_S$$_e$$_t$) to achieve the desired memory window at a current compliance (CC) of 300 nA. The device returns to the HRS during the return path of the voltage sweep at $\approx$ 1 V (V$_R$$_e$$_s$$_e$$_t$). The observed unipolar switching with short-term memory (STM) dynamics is used to create reservoir states, as will be explained in the following section. Excellent uniformity in cycle-cycle (C2C) and device-device (D2D) statistics for the 1L-MoS$_2$ devices can be observed in \textcolor{blue}{Figure S4(b)} as well as in our previous work \cite{R_32}. 
Next, we observe the resistance switching dynamics of the ML-MoS$_2$ devices. The ML-MoS$_2$ memristors require an initial electroforming operation, as shown in (\textcolor{blue}{S5(a)}). Subsequently, they exhibit bipolar and non-volatile dynamics with gradual resistance change in the Set and Reset operations as shown in supplementary \textcolor{blue}{Figure S5(b)}.The quasi-DC \(I-V\) characteristics exhibit smooth analog change in the resistance for different CC values (\textcolor{blue}{S5(c)}). DC endurance measured over 1000 cycles at 600 $\mu$A CC shows outstanding stability and C2C uniformity, as seen in (\textcolor{blue}{S5(d)-(f)}). The histograms of V$_S$$_e$$_t$, V$_R$$_e$$_s$$_e$$_t$, R$_L$$_R$$_S$ and R$_H$$_R$$_S$, measured over 15 devices, can be seen in supplementary \textcolor{blue}{Figure S6(a)-(b)}. \textcolor{blue}{Figure S6(c)} shows stable retention measured at room temperature on a individual cross point ML-MoS$_2$ device. The C2C and D2D statistics of the ML-MoS$_2$ memristors is summarized in Supplementary \textcolor{blue}{Table 1}. Overall,the excellent uniformity and analog resistance switching dynamics of the devices hint at the existence of a non-filamentary mechanism of switching.\\
To develop a clear understanding of the origin(s) of resistance switching, we first note that the ML-MoS$_2$ memristors require an initial electroforming step. This indicates creation of a conductive filament. However, the subsequent switching operation is area-dependent, as shown in Supplementary \textcolor{blue}{Figure S7(b)-(c)}. Therefore, a non-filamentary transport mechanism governs the resistance modulation in the devices after the initial electroforming operation.\\
Next, we perform cross-sectional HRSTEM analysis of the ML-MoS$_2$ under pristine and operational conditions.  The HRSTEM images and corresponding EDS chemical maps of various layers of the device in virgin state are shown in Figure 3a and Supplementary \textcolor{blue}{Figure S8(a-d)}. We observe that ML-MoS$_2$ is a clean layer without any interdiffusion of electrode material(s) into it, with Brown-Powell quantification procedure yielding Mo:S of 1:2 shown in Figure 3(e)(Pristine). 
An operational device imaged after cycling a few times subsequent to electroforming clearly reveals the presence of a filament (Figure 3(b) from HAADF-STEM imaging \textcolor{blue}{Figure S8(e, k)} and corresponding EDS maps (Supplementary \textcolor{blue}{Figure S8(g-j)}). Our quantification procedure in the filament region yields Mo:S of 1:1 revealing an extreme sulfur deficiency (Figure 3(e)(Filament). Thus, it is clear that electroforming involves sulfur vacancy filament formation. 
Furthermore, the non-filament part of MoS$_2$ is also sulfur deficient (Figure 3(c) Figure 3(e)(Non-filament), with the histograms and with some inter-diffusion of Au forming intercalates as can be seen from the HRSTEM image in Supplementary \textcolor{blue}{Figure S8(f))} and Figure 3(d)  (also seen from the chemical maps presented in Supplementary \textcolor{blue}{Figure S9}). These studies paint a clear picture of the switching mechanism in the ML-MoS$_2$ memristors. The switching operation results in the formation of a filament through generation of S-vacancies and Au intercalates in the film. These observations, combined with the analog and area-dependent resistance switching, indicate bulk-limited transport in the ML-MoS$_2$ memristors. Note that there can be an additional interfacial dynamics due to Schottky barrier modulation at the Ti/MoS$_2$ interface. However, this interfacial effect is not dominant at the operating currents used in the non-volatile ML-MoS$_2$ devices. In \textcolor{blue}{Figure 3(f)}, we fit the current-voltage characteristics of the device in HRS and LRS with Space Charge Limited Conduction mechanism. We confirm that the transport in medium and high electric fields is dominated by shallow and deep traps, possibly originating from the Au intercalates and S vacancies. A detailed analysis of the fitting procedure is provided in the Supplementary \textcolor{blue}{Figure S10}. We have previously correlated resistance switching in 1L-MoS$_2$ devices with Schottky barrier modulation\cite{R_32, R_33}. In such devices, the involvement of bulk traps can be ruled out.
\begin{figure}
    \centering
    \includegraphics[width=\linewidth]{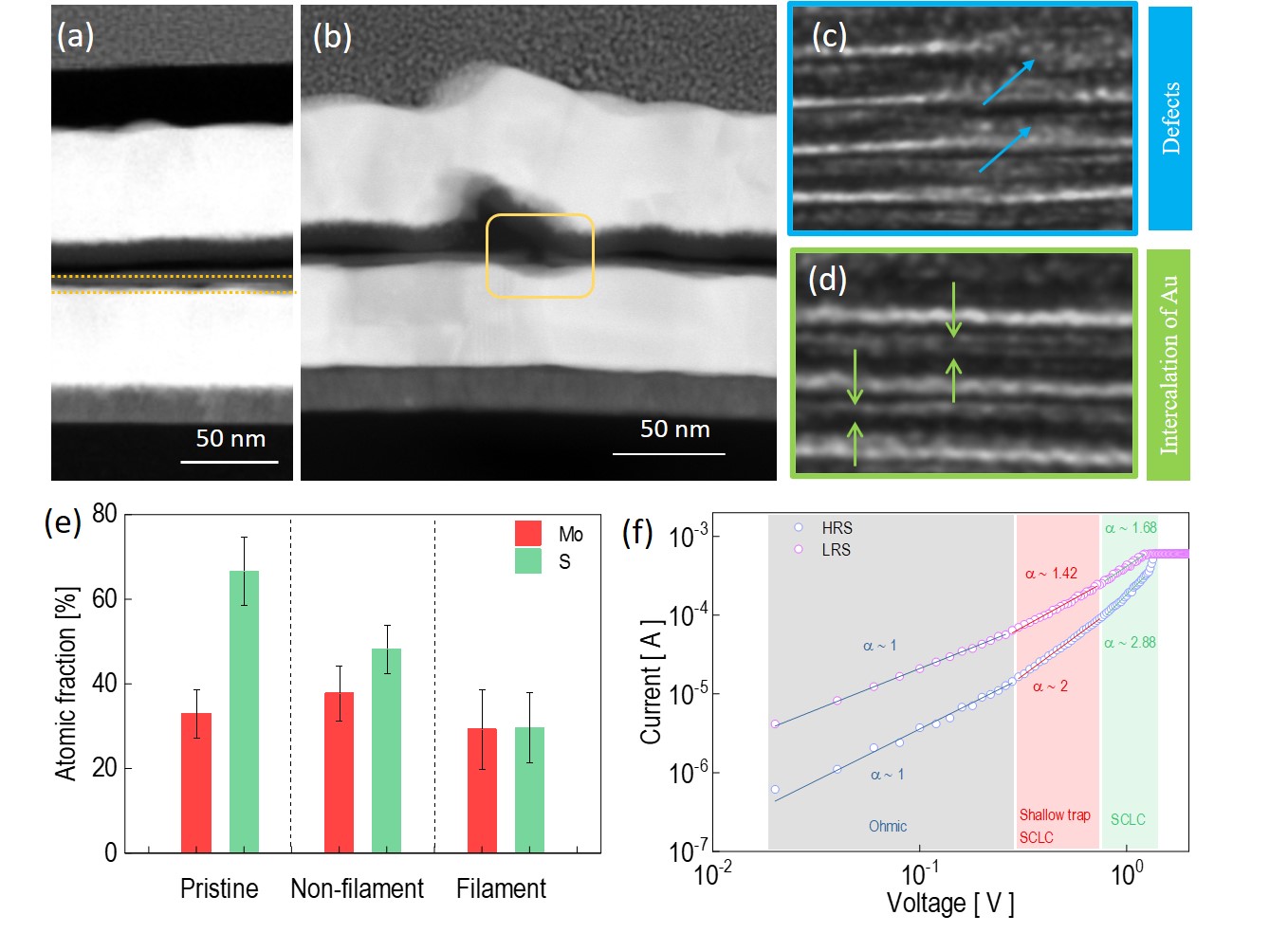}
    \caption{(a) HRSTEM image of a pristine device with the Au/Ti/ML-MoS$_2$/Au stack(from top to bottom). (b) An HRSTEM image of a switched device in the LRS condition showing a conductive path or "filament". (c-d) Defects and Au intercalates in MoS$_2$, away from the filament region. (e) Comparison of atomic fractions of Mo and S in the pristine and switched device shows reduction of S content after switching. (f) Current-voltage characteristics in the HRS and LRS fitted with SCLC transport equation.}
\end{figure}
\subsection{MoS$_2$ memristors as analog synapses and reservoir neurons}
In order to evaluate the applicability of the ML-MoS$_2$ memristors as artificial synapses, we attempt analog conductance modulation through Potentiation/Depression experiments by applying appropriate programming/erase pulses. These tests are carried out for both the individual devices as well as for the 16 x 16 arrays. Conductance modulation in the individual devices is shown in Supplementary Figure S11 and its description can be found in Note 1. C2C variability in P/D operations for the single devices is further evaluated for a large higher number of P/D cycles, as shown in Supplementary \textcolor{blue}{Figure S12}. We also evaluate the C2C variability for long-term P/D operations for different regions of the conductance tuning range. These observations are summarized in Supplementary \textcolor{blue}{Table 2}. We observe that the standard error in conductance has some variation depending on the target conductance. This is expected, given that the conductance tuning characteristic of these devices is not perfectly linear. Despite this, the observed C2C uniformity for P/D operations observed in these devices demonstrate significant improvement compared to most state-of-the-art filamentary memristors\cite{R_37, R_38, R_39, R_40}. Figure 4 summarizes our experiments on the fabricated 16$\times$16 memristor arrays. We have used a floating bias scheme for the program/erase/read operations due to limitations in our existing measurement setup\cite{R_41}. Figure 4(a) shows an SEM image of a 16$\times$16 ML-MoS$_2$ memristor array. The dimension of each crosspoint in the array is 7 $\mu$m. Figure 4(b) shows the pulse-programming scheme used for analog conductance modulation of the devices. A series of identical positive pulses of amplitude 5V and width of 5 $\mu$s are used to continuously increase the conductance from the minimum to the maximum value (potentiation). Conductance is gradually decreased by applying a series of identical, 5 $\mu$s pulses with -2V amplitude (depression). Figure 4(c) exhibits the potentiation/depression operation performed over many cycles on a particular device in the array. Figure 4(d) shows analog conductance modulation performed on multiple devices in the array. Note that the sneak-current in a passive crossbar array can lead to some device-to-device (D2D) fluctuations. Long-term endurance of the devices is evaluated up to 10,000 cycles, as shown in Figure 4(e). The endurance measurement uses pulse amplitudes of +2V, -3.5V and 0.5V for program, erase and read operations, respectively. No degradation of memory window is observed for the device. Finally, long-term potentiation/depression operations over more than 5000 cycles are performed to test the stability of the devices under stress (Figure 4(f)). The cumulative probability resulting from the measured conductance at every 25$^{th}$ pulse of every cycle is plotted in Figure 4(g). We observe a standard error less than 4$\%$ in conductance modulation from these experiments.The variability observed in filamentary memristors is typically higher. Additionally, most prior works on filamentary memristors report C2C fluctuations observed from quasi-DC switching. The variability in conductance tuning for filamentary devices is expected to be even higher than those obtained under quasi-DC programming conditions, due to shorter programming times involved. In summary, the bulk switching ML-MoS$_2$ memristors provide considerable benefits over filamentary memristors in terms of switching uniformity. 
\begin{figure}
    \centering
    \includegraphics[width=\linewidth]{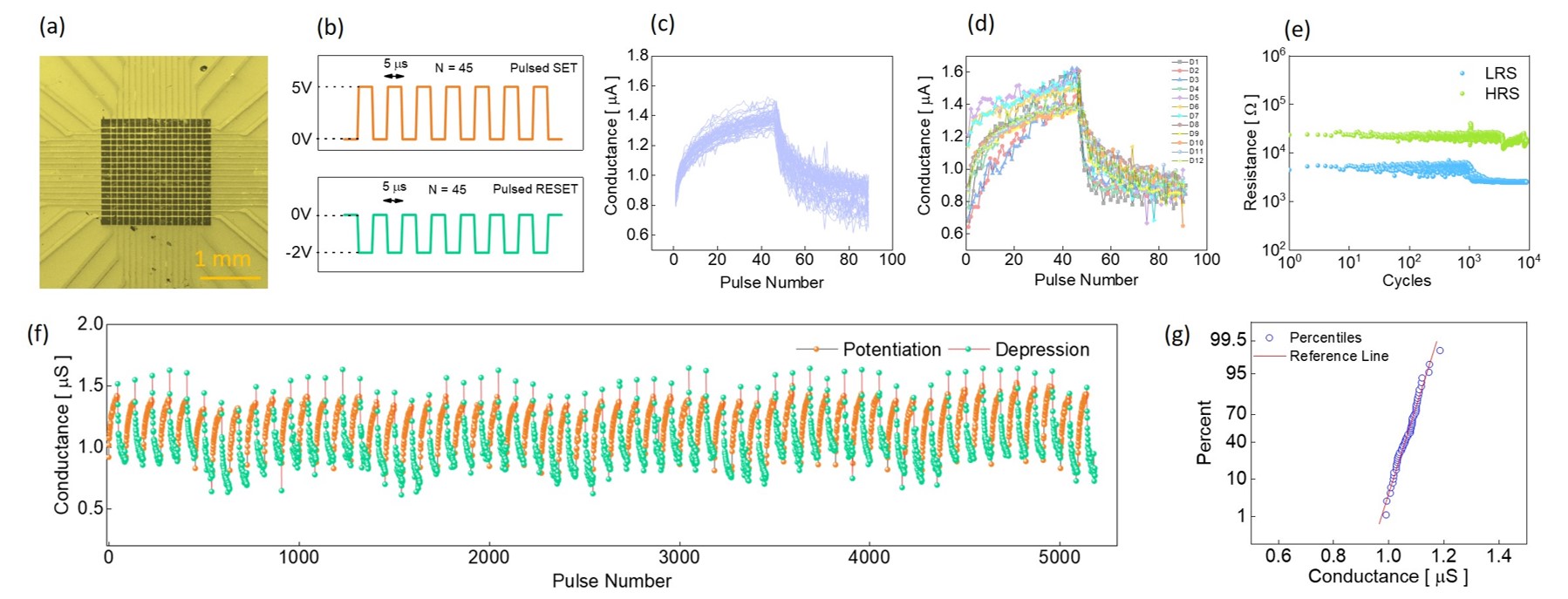}
    \caption{Characteristics of 16$\times$16 ML-MoS$_2$ crossbar array. All the devices in the array have an area of 7 $\mu$m $\times$ 7 $\mu$m. (a) An SEM image of the developed array, (b) Pulse programming scheme for potentiation (P) and depression (D) operations. One complete P/D cycle starts with potentiation using 45 identical positive pulses followed by depression with 45 identical negative pulses. (c) Multiple P/D cycles performed on one device in the array. (d) P/D operation performed on different devices in the array. (e) Measured endurance up to 10,000 cycles. (f) P/D operation performed over more than 5000 pulses on a device, demonstrating long-term stability of conductance modulation. (g) Cumulative probability of device conductance at 25$^{th}$ pulse in each potentiation cycle.}
\end{figure}
\\Next, we study the short-term memory (STM) characteristics of the Au/Ti/1L-MoS$_2$/Au devices to evaluate their usefulness as reservoir neurons. STM can be studied by applying a 4-bit input pulse stream to the Au/Ti/1L-MoS$_2$/Au devices. We vary the input pulse stream from creating states from [0000] to [1111]. The input pulse configuration for creating the state [1111] is shown as an example in Figure 5(a). initially, a pulse of 2V is applied to read the initial state of the device followed by a 4-bit configuration pulse of 6 V programming and 2 V read with pulse width of 10 $\mu$s. We observe that the conductance of the device continues to increase if consecutive programming pulses are applied. On the other hand, conductance will decay in the absence of a programming pulse.  In this manner, 16 possible states can be achieved ranging from [0000] to [1111], as shown in Figure 5(b). This dynamic behavior can be used to emulate the "reservoir" states. Figure 5(c-f) shows the measured response of 4 different input signals [0010], [0101], [0110], and [1010], repeated more than 15 times on a same cell to check the cycle-to-cycle variation. Each response from the input signal gives the unique output to be able to distinguish between each pattern of the input. The variation of 16 possible states measured on various devices has been shown in Supplementary \textcolor{blue}{Figure S13}.\\
\begin{figure}
    \centering
    \includegraphics[width=\linewidth]{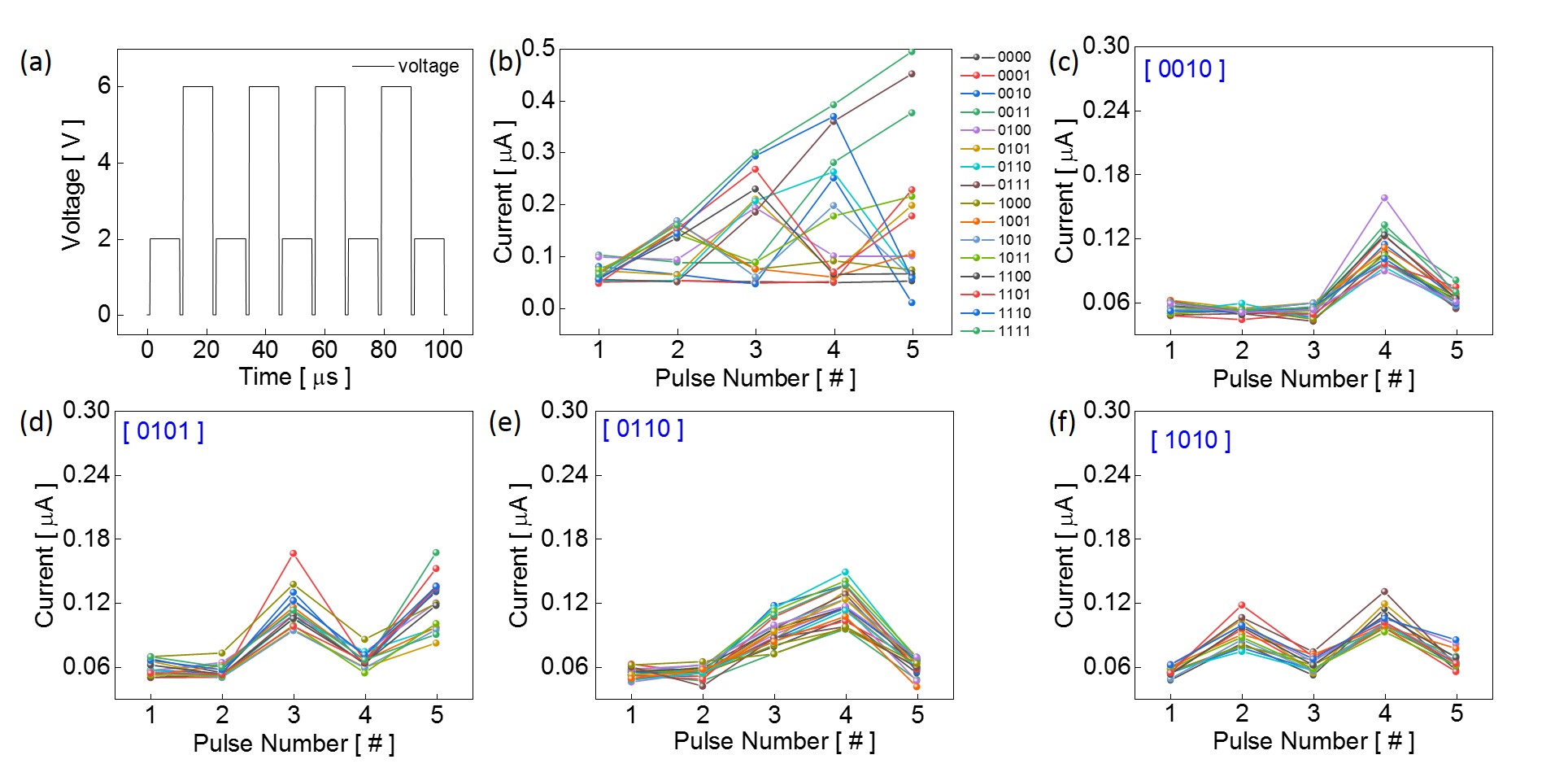}
    \caption{(a) The input pulse stream to create [ 1111 ] state. (b) 16 output states for the input pulse stream varying from [ 0000 ] to [ 1111 ]. Response of a device repeated for same signal recorded 16 times for pulse streams (c) [0010], (d) [0101], (e) [0110], (f) [1010], (Device area 7 $\mu$m $\times$ 7 $\mu$m).}
\end{figure}\\
Finally, we design and simulate a reservoir computing framework by leveraging the rich temporal dynamics of 1L-MoS$_2$ memristors and the synaptic functionalities of a 16$\times$16 ML-MoS$_2$ memristor array. The designed RC framework is used to demonstrate spoken-digit recognition and prediction of a time-series equation.

\subsection{Realization of RC framework for spoken digit recognition}
The implementation of the framework begins with the experimental recording of the reservoir states, using the volatile memory behavior inherent to 1L-MoS$_2$ resistive switching devices \textcolor{blue}{(Figure S4(a))}. This is achieved by applying a specific pulse scheme to the device: write1-read1-write2-read2-write3-read3-write4-read4. By modulating the bias amplitudes for the write1, write2, write3, and write4 operations between high (1) and low (0), and reading the device’s current or resistance after each write operation, we can extract up to 16 distinct states from a single device (Figure 5(b)). For example, with a [0000] scheme, where a low write bias amplitude is applied for each write pulse, the recorded device current is low for all four read pulses, as none of the individual write pulses is capable of turning on the device. In contrast, with a [1000] scheme, the device exhibits a relatively high current during the initial read pulse, as it is stimulated with a high write bias amplitude. However, the short-term memory effect causes the device's resistance to increase over time, as there is no significant disturbance applied during the next three write pulses with low bias amplitude. The current in the device will decrease with time, as observed during the application of the remaining three read pulses. Hence, using the defined scheme, we have recorded 16 different states for each device, with each state having four recorded read currents, as shown in Figure 5(b) and supplementary \textcolor{blue}{Figure S13}.
In addition to expected device-to-device variation, all devices have given the expected response to all input stream variation by following a similar current dynamics.
\\We have used the Free Spoken Digit Dataset (FSDD), which contains waveforms of digits (0-9), spoken by various speakers. In the dataset each waveform is sampled at 8 kHz and padded or truncated to 16,000 samples. The input waveforms from the Free Spoken Digit Dataset are first pre-processed to extract the thirteen-dimensional Mel-Frequency Cepstral Coefficients (MFCCs) per frame [\textbf{Supplementary Information}]. Following this, the pre-processed data features are fed into the masks connected to the eight reservoir nodes (represented by the STM resistive switching devices).\\ 
These masks are responsible for encoding the incoming data into four-bit sequences composed of 0's and 1's. The reason for using the specified encoding scheme is the experimental recording of 16 distinct states for each of the five devices. We use a "look-up table" in the code, containing normalized read current values, to map the relationship between the input bit stream and the corresponding device response. The outputs from the reservoir nodes are fed to the output layer forming the "read-out layer", analogous to a conventional feed-forward artificial neural network (ANN). During simulation, the read-out layer weights are trained using the training dataset. The schematic representation of this system is given in Figure 6(a) which consist of eight reservoir neurons and two hidden readout layer. Figure 6(b) represents the streams of 0's and 1's converted from the audio input. Figures 6 (c) and 6 (d) present a confusion matrix and the variation in accuracy with epoch for the classification of the test dataset.\\
Non-volatile resistive switching devices have been studied as excellent choice for hardware network weights, provided they are capable of smooth and gradual conductance tuning. Gradual conductance tuning, in contrast to abrupt tuning, supports effective learning of features and weight training. The gradual conductance-tuning property of the nonvolatile devices is utilized to prepare the weights in the readout layer. The use of devices based on 1L and ML MoS$_2$ for the reservoir layer and the readout layer, respectively, is also motivated by the fact that it's beneficial to build the entire RC network on the same material platform. The weights are updated using the Manhattan update rule, which generates the new weight value by adding the product of the learning rate and the sign of the gradient of the loss function to the old weight value. The learning rate is mapped to the device's conductance-tuning behavior on the basis of the sign of the gradient of the loss function. If the sign is positive, the learning rate is equal to the difference between the final and initial conductance, divided by the number of pulses required to transition between the two. The potentiation behavior of the device is studied for the following computation. For the negative case, the learning rate is derived from the device's depression behavior. In addition to the mentioned weight-update rule, the effect of experimental cycle-to-cycle deviation is also included by adding a noise factor in the calculation of the learning rate.  Although a simplified approach is used to map the nonlinear device behavior to the weights, MoS$_2$-based nonlinear devices could be an excellent choice in practical hardware implementations to offer gradual modulation over a broader conductance range. While several prior reports exist on reservoir computing with memristive technologies \cite{newrev2,R_25,newrev3}, , our work demonstrates a memristive RC on a single 2D-material platform, where both the reservoir layer and the readout layer are realized using MoS$_2$ memristive devices. The simulated results show a word error rate (WER) of around 0.145 percent, which may increase in hardware demonstrations due to practical non-idealities, but is still expected to remain within the 0.2–0.4 percent range, consistent with state-of-the-art results \cite{R_10}. The effect of noisy input is also explored by injecting Gaussian noise into the input dataset (Figure  6(d)). As expected, the classification accuracy decreases with increasing noise intensity, but still remains within an acceptable range. However, the distortion introduced by noise is strongly dependent on the noise frequency relative to the input waveform, which in turn decides the capability of the volatile devices in the reservoir layer to forget the noise perturbations.
\begin{figure}
    \centering
    \includegraphics[width=\linewidth]{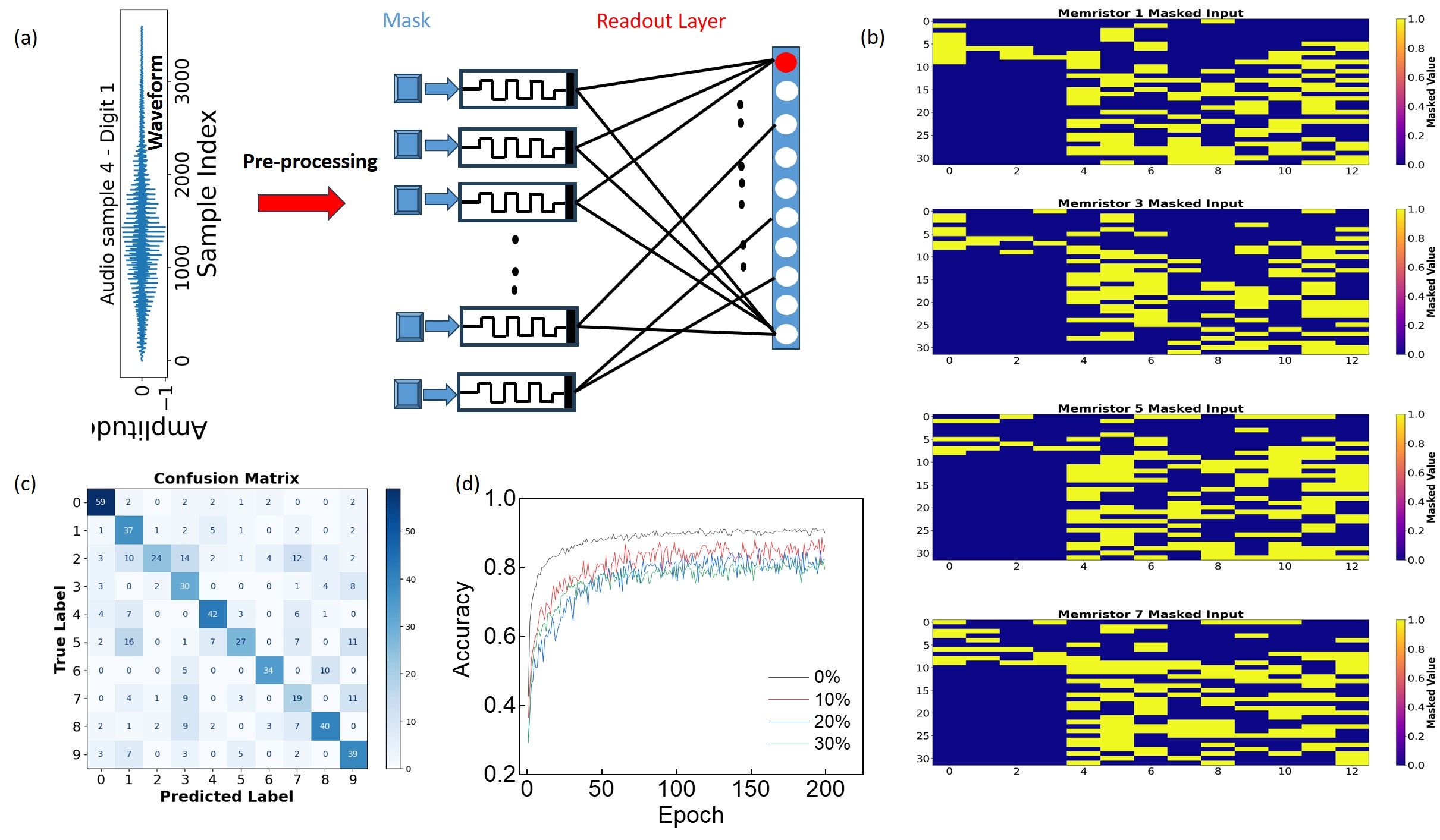}
    \caption{(a)  A reservoir computing framework for spoken digit recognition consisting of pulse input layer, reservoir layer with volatile devices (STM), and readout layer consist of two hidden layer of nonvolatile devices (LTM), (b) Conversion of audio signal to streams of 0's and 1's for input, (c) Output of a confusion matrix, (d) Accuracy vs epoch for a design with STM devices at reservoir layer and readout layer LTM weight updates inspired by conductance behaviour observed for devices in a 16 × 16 array.}
\end{figure}
\subsection{Solving non-linear dynamic time-dependent equation}
The dynamics of complex electrical systems involves solving non-linear equations, to understand and optimize system control. Reservoir frameworks could be used to address such time-dependent nonlinear equations. To demonstrate the use of a memristor-based reservoir system for this task, we consider the following equation:
\[y[k]=0.1y[k-1]+ 0.2y[k-2]y[k-3]+0.3u[k]^3+ 0.25\]
The equation is a modified version of an equation previously discussed in the literature (Atiya et al. \cite{rf-ts1}  ) for recurrent neural network training. The output at time frame k depends on the input at time frame "k" and on the product of outputs at previous time frames, i.e. "k-2" and "k-3". The input data are normalized to lie within the range [0, 1]. The normalized data are then mapped to a stream of four bits, ranging from 0000 to 1111. 
The reservoir network \textcolor{blue}{(Figure 7(a))} consists of five memristors, the input layer comprising five input nodes: u[k],..,u[k+4]. Each input node is connected to a mapping block, which converts the input data into a 4-bit binary stream. The mapping block responses are then applied to the memristors. For all possible cases of 4-bit streams (0000 to 1111), the corresponding memristor outputs are pre-stored in a lookup table. The final state of the memristor, determined by the applied bit stream, serves as input to the readout layer, which computes the final output y [k] as shown in \textcolor{blue}{Figure 7(b)}. The readout layer, in this case, consists of two hidden layers with 128 and 64 neurons, respectively. We observe excellent agreement between the true values and the predicted values of the time-series equation. However, it is important to note that in this case, the learning method is offline, where the model is trained on the dataset in batches and weight updates occur only after processing each batch. Offline learning could face a challenge in adapting to real-time changes in the data, motivating the use of online learning, where the model updates its weights after each individual input, enabling real-time adaptation. In online learning, the network also has the capability to continuously update its weights after processing each individual input, leveraging immediate feedback from the prediction error. Figure 7(c) demonstrates the true value vs predicted value comparison when the model leverages online learning.

\begin{figure}
    \centering
    \includegraphics[width=\linewidth]{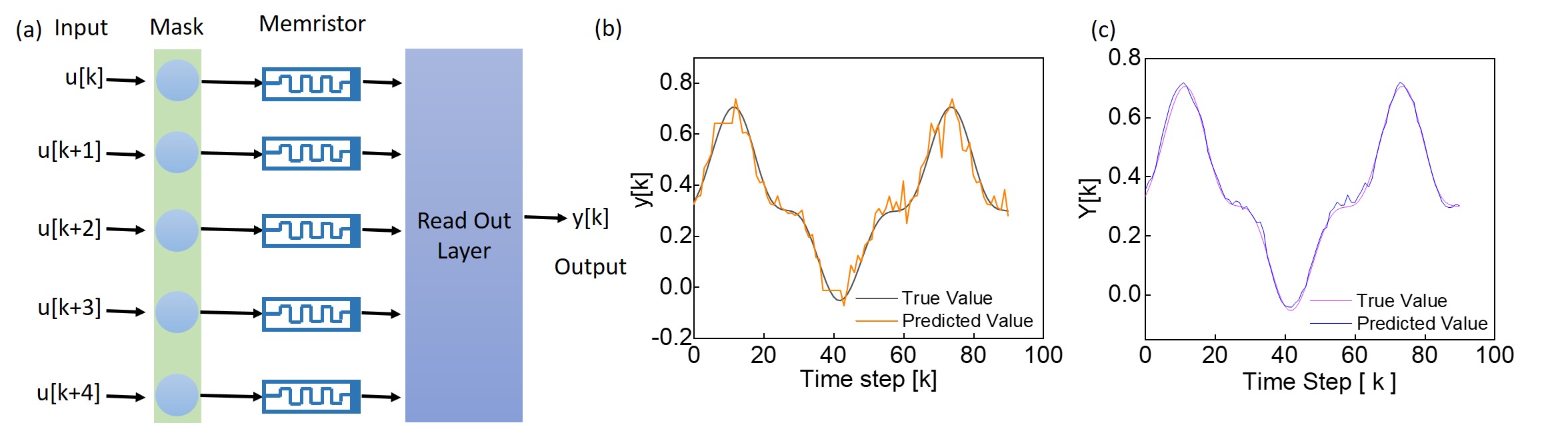}
    \caption{(a)  A reservoir computing framework for predicting output of a non-linear dynamic time series equation. (b) Prediction of y[k] using memristor based neural network [Offline Learning], (c) Prediction of y[k] using memristor based neural network [Online Learning].}
\end{figure}

\section{Analysis of energy cost}
The current work has been compared with some recent advancements in memristor-based RC frameworks (Table 3, Supplementary Information).
In both applications, a masking layer is employed to convert data from its real-valued form to a stream of bits, which serves as input for memristive devices. The hardware implementation of the mapping block could utilize analog-to-digital converters (ADC) and logic gates. The energy requirement for an ADC is determined by the product of the load capacitance, the number of output levels, and the operating voltage. With four output levels and operating voltage dependent on the ADC specifications, optimizing the load capacitance can result in energy consumption on the order of a few picojoules. The energy consumption for the logic gates is minimal compared to that of the ADC and the energy cost for the mapping block could be closely related to only that of the ADC. Additional energy requirement comes from the memristive devices used in the reservoir layer, calculated based on the amplitude, pulse width, and current response of the devices. The reported devices operate with input pulse stimulation time in the microsecond range and exhibit current responses in the range of hundreds of nA. As a result, energy consumption for the reservoir layer is also in pJs. Further optimization of the device performance could lead to reductions in stimulation amplitude and stimulation time needed for device response. Another key energy factor arises from the readout layer. In hardware, this layer is intended to be implemented using long-term memory devices, which also respond to pulse stimuli with on-times in the microsecond range. Optimizing the pulse stimulation time in both the reservoir and the readout layer depends on factors such as the experimental setup and device dynamics. We use pulse stimulation time in the microsecond range to ensure device current stabilization, enabling accurate state readings.\\
To estimate the efficiency of the network, we focus on two parameters- the number of operations and the power consumed. For the first application, the number of training operations required is around 150 and the efficiency is estimated to be around 181,818,182 OPS/W. Please note that the epoch time for this case will be roughly around 5.5 ms if 1 s is assumed as the epoch time for the other case. The number of operations involved in training the readout layer for one epoch for the second application is calculated to be around 27200. For a fully memristive readout layer, 8,896 memristors are required and assuming a single epoch time of 1 second, the total power consumption for this setup would be around 8,896 \si{\micro W}. The efficiency of the network for the second application is estimated to be approximately 3,057,553 OPS/W (Operations per Second per Watt).

\section{Conclusion}
In conclusion, we have successfully demonstrated the "spoken digit recognition" and "time dependent nonlinear equation solving" reservoir computing system by using short-term memory and long-term memory devises fabricated with CVD grown large area monolayer and multilayer MoS$_2$ 2D material. By exploiting the STM effect of 1L-MoS$_2$ based devices, we successfully extracted the 4-bit code with 16 possible states for RC computing. The LTP and LTD data have also been extracted from the ML-MoS$_2$ based devices which can be used for training purposes in the neural network.
\section{Experimental Section}
\subsection{Material}
MoO$_3$ (Molybdenum Trioxide) (99.5$\%$ ARIACS) powder and S (Sulfur) (99.5$\%$ AR) powder are purchased from LOBA CHEMIE PVT.LTD. and used to synthesize the CVD grown monolayer and multilayer MoS$_2$ continuous film on SiO$_2$/Si substrate having SiO$_2$ 300 nm thick, 1-10 $\Omega$-cm resistivity with orientation (100).

\subsection{CVD Growth of Monolayer and Multilayer MoS$_2$}
We used a single-zone furnace for the CVD growth of MoS$_2$, which is divided into two sections by separate heaters. The system is organized as follows, according to \textcolor{blue}{Figure S1}: (a) the main split chamber (reaction chamber) where the substrate is positioned alongside the primary precursor, Molybdenum (MoO$_3$); (b) the pre-heater chamber where the secondary precursor, Sulfur (S), is located; (c) the carrier gas flow inlet and outlet; (d) a mass flow controller (MFC) for regulating the gas flow rate; and (e) PID controllers for managing the temperatures of the main furnace and pre-heater, respectively.\\
Sulfur powder and MoO$_3$ powder are placed in designated boats. A clean Si/SiO$_2$ sample with a 300 nm thick SiO$_2$ layer is placed on the boat containing the Molybdenum precursor, with the sample's top surface facing the Molybdenum (upside down). Then both boats are inserted into tube 1, which is placed inside tube 2 of the main system, as illustrated in \textcolor{blue}{Figure S1}. The distance between the two precursor sources is maintained at 40 cm.\\
The system is purged at a flow rate of 200 SCCM for 3 to 4 cycles to remove all oxygen and prevent unwanted reactions. The temperatures of the main chamber (furnace) and preheater are controlled using PID controllers in three phases: rise time to reach the target temperature, hold time for the reaction, and cooling time to lower the temperature. The MFC regulates the carrier gas flow rate, which transports the precursors in their gaseous forms to the substrate. Either N$_2$ or Ar can be used as the carrier gas. Various process parameters have been tested to optimize the growth of monolayer and multilayer MoS$_2$, with the temperature profiles for each presented in \textcolor{blue}{Figure S2}.


\subsection{Device Fabrication}
The fabrication of the two-terminal memory device began with cleaning the SiO$_2$/Si sample using TCE, acetone, and IPA. This was followed by patterning the bottom electrode (BE) through photolithography, depositing 10 nm of Cr and 40 nm of Au via e-beam evaporation, and performing a lift-off process.
Next, CVD-grown 1L and multilayer MoS$_2$ were transferred to separate samples using the chemical transfer method. In this method, MoS$_2$ was coated with a PMMA solution and heated on a hot plate at 100 $^0$C for 5 minutes. The PMMA-coated sample was then immersed in a KOH solution, which removes the SiO$_2$ and causes the MoS$_2$/PMMA stack to float. Once the MoS$_2$/PMMA stack was fully detached, it was transferred to deionized water (DI) to remove residual KOH. The MoS$_2$/PMMA stack was then transferred to a substrate patterned with Cr/Au BE and left in a desiccator overnight. Subsequently, the sample was heated on a hot plate for 10 minutes to remove any remaining DI water. The PMMA layer was then removed using acetone, followed by DI water rinse and dehydration.
Finally, the transferred MoS$_2$ was patterned using photolithography and an etching process. The top electrode (TE) was fabricated in the same manner as the bottom electrode, with 10 nm Ti and 40 nm of Au deposited via e-beam evaporation and a lift-off process. Some of the optical and SEM images of devices given in Supplementary \textcolor{blue}{Figure S14}

\subsection{Characterization}
The as-grown MoS$_2$ samples are explored in an optical microscope (Nikon, model no. ECLIPSE LV100ND) under a 100× objective lens with NA ~0.95.
The Raman and PL measurements are performed at RT using the micro-Raman system (Renishaw Invia spectrometer) using 532 nm laser excitation with gratings of 2400
grooves/mm (Raman) and 600 grooves/mm (PL), respectively.
The laser power is kept constant throughout the measurements.
AFM measurements were performed on a Park NX10 system in noncontact mode at room temperature.
A lamella was prepared for detailed structural analysis using a focused ion beam (FIB) (Model FIB-Helios 5).
The monolayer and multilayer stacking of MoS$_2$ was examined using Titan Themis 300 at 300 kV and compositional analysis was also performed on the same instrument for high precision elemental characterization.
The quasi-DC measurements are analyzed using a B1500A parametric analyzer.
Pulse measurements are analyzed using the WGFMU B1500A waveform generator.
Python has been used for simulation. 
\begin{acknowledgement}

This work is partially funded by the Science and Engineering Research Board (SERB), a statutory body of the Department of Science and Technology, Government of India, through the project SRG / 2020/001126 and by the Research Center for 2D research and innovation at the Indian Institute of Technology Madras through the Institute
of Eminence (IoE) Scheme.
The authors also acknowledge the Micro and Nano Characterization Facility (MNCF) at CeNSE, IISc Bengaluru, India, for STEM sample preparation and STEM measurements.
\end{acknowledgement}

\begin{suppinfo}
\begin{itemize}
    \item Schematic diagram of CVD system, temperature profile.
    \item SEM images of CVD grown monolayer and multilayer MoS$_2$ material.
    \item Analysis on volatile and non-volatile memory devices.
    \item 16-output states measurements done on five different devices.
    \item Optical and SEM images of the fabricated devices.
 
\end{itemize}
\end{suppinfo}
\bibliography{achemso-demo}


\end{document}